\documentclass[epj]{svjour}
\usepackage[centertags]{amsmath}
\usepackage{epsfig}

\def\openone{{\leavevmode\hbox{\small1\kern-3.55pt\normalsize1}}}

\newcommand{\e}{\langle e \rangle}

\begin{document}

\title{Slow dynamics due to entropic barriers in the one-dimensional 'descent model'}
\author{V. DESOUTTER \and N. DESTAINVILLE}
\institute{Laboratoire de Physique Th\'eorique, IRSAMC - UMR CNRS/UPS 5152 \\
Universit\'e Paul Sabatier,
118, route de Narbonne, 31062 Toulouse Cedex 04, France.\\
}

\abstract{
We propose a novel one-dimensional simple model without disorder
exhibiting slow dynamics and aging at the zero temperature limit. This
slow dynamics is due to entropic barriers. We exactly solve the
statics of the model. We derive an evolution equation for the slow
modes of the dynamics which are responsible for the aging. This
equation is equivalent to a random walker on the energetic landscape. This
latter elementary model can be solved analytically up to some basic approximations
and is shown to present aging by itself, as well as a slow logarithmic relaxation
of the energy: $\langle e\rangle(t) \sim 1/\ln(t)$ at large $t$.
} 

\titlerunning{Slow dynamics in the one-dimensional 'descent model'}

\maketitle

\section{Introduction}

Below their glassy transition temperature, glassy systems relax very
slowly and remain out of equilibrium on experimental time
scales. Despite intensive theoretical studies, the exact nature of
this transition is not clearly understood yet, even if it is widely
believed that it is of dynamical nature \cite{h.Walterkob_02}.
Glassy systems seem to be trapped in some metastable states whose numbers 
and life times increase dramatically with decreasing energy.
Therefore there is an increasing demand for paradigmatic toy-models
containing the elementary physical mechanisms responsible for
glassiness and its experimental manifestations, such as the
aging effect \cite{b.Bouchaud_al_97}.
Among several other mechanisms for slow relaxation and aging, the focus
has recently been brought into the existence of entropic barriers
\cite{r.godreche_1995}. This term designates boundaries between regions of the phase 
space where the system has very rare possibility to find a path to go
from one region to the other. The system can be trapped in such a region
without the necessity of energetic barriers. By analogy with Arrhenius
law, the height $\Delta S$ of an entropic barrier is defined by
$\tau=\exp(\Delta S)$, where $\tau$ is the time the system needs to go
through the entropic barrier. Simple models~\cite{Ritort_02}~-- such as the {\em
backgammon} model \cite{r.ritort_95} or {\em urn} models
\cite{r.Godreche_01}~-- have been developed which exhibit such
entropic barriers.  These models are of mean-field type such as most
of models for glassy dynamics.  As compared to these simple models for
glassiness and entropic barriers, the interaction in the present toy-model is
one-dimensional and therefore it can be legitimately considered as more physical. 
 
On the other hand, our model is based on the
well-characterized permutation group $\Sigma_n$, and many calculations
can be carried out either exactly or after some basic assumptions.

\section{The model}

We consider a one-dimensional system of size $n$ where the
configurations are the $n!$ permutations on $\{1,\dots,n\}$ of the
symmetric group $\Sigma_n$. We represent a configuration $\sigma \in
\Sigma_n$ by a {\itshape word}
$\sigma=\sigma(1)\sigma(2)\dots\sigma(n)$ with $\sigma(i) \in
\{1,\dots,n\}$.  The energy $E$ of a configuration is defined by its
number of descents. One says there is a descent between $i$ and $i+1$
if $\sigma(i)>\sigma(i+1)$.  For example, let $n=6$, the permutation
\begin{equation}
\sigma_{ex}=15\!\downarrow\!236\!\downarrow\!4
\end{equation}
has 2 descents symbolized by an arrow $\downarrow$ and its energy is
therefore $E(\sigma_{ex})=2$.  

We denote by $D_n^k$ the degeneracy of the energy level $E=k$. The
$D_n^k$ are known as Euler numbers\cite{euler_1755}. The identity is
the unique ground state with zero energy so that
$D_n^0=1$. Furthermore we can define a symmetric permutation
$\sigma^c$ for all $\sigma \in \Sigma_n$, by
\begin{equation}
\sigma^c=n+1-\sigma(i), 
\end{equation}
for $1\le i \le n$, which has the energy $E(\sigma^c)= n-1-E(\sigma)$.
The distribution of configurations with energy $k$ is symmetric with
respect to $(n-1)/2$, and $D_n^k=D_n^{n-1-k}$. At high temperatures
all configurations have the same probability, the mean energy per
particle in this limit is $\langle e \rangle = (n-1)/2n$, according to
the above symmetry and $\langle e \rangle = 1/2$ at the large
size limit.

The system evolves {\it via} a Metropolis Monte-Carlo algorithm
\cite{b.newman} and we study two different dynamics: in the non-local
one, at each time step, any two sites $i\neq j$ are chosen at random
and we try to transpose the elements $\sigma(i), \sigma(j)$ with a
certain probability $P_{ji}$, which depends on the temperature $T$ and
on the energy variation $\Delta E$ if the transposition were executed:
$P_{ji}=\min(1,\exp(-\Delta E/T))$. In the local dynamics we choose
only one site $i$ at random, and we try to transpose $\sigma(i)$ and
$\sigma(i+1)$, with the same transition probability as
above. Sometimes in the following we will call {\it particles} the
$\sigma(i)$. Indeed, this model can be seen as a system of $n$
distinguishable labelled particles on a one-dimensional lattice which
tend to sort themselves.

We anticipate on the following to emphasize the origin of the entropic
barriers in the non-local dynamics. If {\it quenched} from infinite to
vanishing temperature the system evacuates its energy and relaxes
through decreasing energy levels towards its 
equilibrium state. We shall demonstrate in the following that it
encounters entropic barriers between any two successive energy levels
whereas it relaxes rapidly inside each energy level. For example let
us explain why there is an entropic barrier between the first excited
states and the unique fundamental one.  We need to know more precisely
the values of $D_n^k$. Starting from the expression \cite{euler_1755}:

\begin{equation}
(k+1)^n=\sum_{j=0}^{k}D_n^k\binom{n+k-j}{n} \label{equ_k^n}, 
\end{equation}
we derive the value of $D_n^1=2^n-(n+1) \simeq 2^n$ for $n \gg 1$.
We get in the same way:
\begin{equation}
D_n^k \simeq (k+1)^n, \label{equ_Dnk} \mbox{ for } k/n \ll 1.
\end{equation}
We can easily recover the result for $D_n^1$ by constructing the
permutations with one descent as follows: we place the elements of the
permutation, beginning by $1$ and continuing with $2$ {\it etc...},
randomly in two different sets, like if we were distributing
distinguishable particles into two boxes. We can build $2^n$
permutations that have a descent at the boundary of the two sets,
except if we have built the identity, which is obtained in $n+1$ ways,
moving this boundary between any two successive elements of the
identity. Subsequently $D_n^1=2^n-(n+1)$. This point of view can be
generalized to any energy $k$. The system is seen as $(k+1)$ ordered
subsets separated by descents. In this point of view, the descents can
be considered as domain walls between ordered subsets.

This point of view also highlights the similitude between our model
and the {\it backgammon} \cite{r.ritort_95,r.franz_ritort_96} one. It
consists in $N$ distinguishable particles placed in $N$ boxes, where
the energy is equal to minus the number of empty boxes. Our subsets
play the role of the boxes in the backgammon model. The difference
with our model is that the particle interaction in the backgammon
model is not localized: it is intrinsically mean-field. Note however
that there exists a one-dimensional generalization of the backgammon
model~\cite{crisanti_00} where particles are allowed to jump to
nearest neighbors only. However, this generalization complicates the
analysis of the model~\cite{crisanti_00}.

We have seen above that there are about $2^n$ first excited states and
only one fundamental. In order to find this state, the system with
energy equal to one, will wander in the phase space until it finds a
state connected to the fundamental one. If the wandering is uniform in
the first level, the probability that the system finds such a state,
will be $(n(n-1)/2)/2^n$, where $n(n-1)/2$ is the number of path
starting from the fundamental state in the non-local dynamics. Being in one of these states, it
will have a probability $2/n(n-1)$ to choose the direction towards the
minimal energy.  So, there will be about $2^n$ steps of the dynamics
before the system finds the identity state, which corresponds to an
entropic barrier of height of order $n$. Note that the previous argument also
holds in the local dynamics: one obtains the same entropic barrier.
Nevertheless, the type of the dynamics will be of
importance, since one can suppose an uniform wandering only for the
non-local one for which we can prove the connectivity of the energy
levels even at $T=0$. Which means that there is always a path between any two
configurations of one level that is inside it. The system
searching the paths to decrease its energy will not have to pass
through any energy barrier. It is not the case for the local dynamics.
But the entropic barriers, which depend mainly on the ratio of the
level sizes, will be also present in the local dynamics, in conjunction
with energetic barriers. In terms of permutation, passing through an
entropic barrier corresponds to a complex rearrangement
of particle. The order in which pairs of particles must be
swapped to optimize their position without increasing the energy
is quite constrained.

\section{Statics}

One of the advantages of the model is that its partition functions,
canonical or grand canonical, have been extensively studied in the
mathematical literature.
Indeed, we can write the canonical partition function as follows:
\begin{equation}\label{equ_zdef}
Z=\sum_{k=0}^{n-1}D_n^k \exp\left(-\frac{k}{T}\right)=
  \sum_{k=0}^{n-1}D_n^k t^k 
    = \frac{1}{t}(1-t)^{n+1}\sum_{k\ge 0}k^nt^k,
\end{equation} 
with $t=\exp(-1/T)$ and where the last equality comes from combinatorial analysis \cite{b.comtet2}.
More precisely, it is obtained inductively using the following relation on the $D_n^k$:
\begin{equation}\label{equ_indDnk}
D_{n+1}^k=(n+1-k)D_n^{k-1}+(k+1)D_n^{k}.
\end{equation} 
This former equation is obtained by building configurations with $n+1$
elements by adding the $(n+1)^{th}$ particle in all the possible
positions on a configuration with $n$ elements. One can further
remark that Eq.~(\ref{equ_Dnk}) comes from the second term of the
right hand-side of Eq.~(\ref{equ_indDnk}).  To obtain the mean values
of the thermodynamical observables we compute:
\begin{equation}\label{equ_lnZ}
\frac{\ln Z}{n}=\frac{n+1}{n}\ln(1-t)+\frac{1}{n}\ln\sum_{k\ge 0}k^nt^{k-1}.
\end{equation}
The sum in the right hand side of the former equation reads
\begin{eqnarray}
\sum_{k\ge 0}k^nt^{k-1} & = &
\sum_{k \ge 0}\exp-n(\ln(k+1)-\frac{k}{nT}) \\ \nonumber
 & = & \sum_{k\ge 0}\exp(-nf(k,T)) \\ \nonumber
 & = & ( 2\pi nT^2)^{1/2}\exp(1/T-n),
\end{eqnarray}
by taking the continuous limit, and evaluating the so-obtained integral by a saddle-point argument for $n
\gg 1$. This continuous limit is only valid if the width of the Gaussian approximate of
$\exp\left(-nf(k,T)\right)$, is large as compared to the spacing
between two consecutive energy levels, $\Delta k=1$, that is to say if
$T \gg 1/2\sqrt{n}$. In the thermodynamic limit it will be always true
for any finite temperature. Using
Eq.~(\ref{equ_lnZ}) and thermostatics identities the
thermodynamical observables can be exactly computed. The free energy per
site obeys 
\begin{equation}
f=-T(\ln(1-t)+\ln(nT)) 
\end{equation}
and the mean energy per particle
is given by 
\begin{equation}
\langle e\rangle=T-t/(1-t). 
\end{equation}
It follows that the specific heat is
\begin{equation}
C_V=1+t/(T(1-t))^2, 
\end{equation}
and the entropy per particle is given by
\begin{equation}
s=(\langle e\rangle-f)/T=\ln(nT)+\ln(1-t)-t/T(1-t)+1. 
\end{equation}
Note that this entropy is not extensive. 
Let us remark the existence of an interesting low temperature domain
defined typically by $T<0.1$, where the first term of
Eq.~(\ref{equ_lnZ}) is negligible so that $\langle e\rangle
\simeq T$. It delimits a low energy domain,
where $D_n^k \simeq (k+1)^n$, which results in entropic barriers. We
will see in the following part that the slow dynamics takes place in
this region of the configuration space.
These results show that there is no thermodynamical
transition at any finite temperature.

\section{Dynamics}

We have defined two types of dynamics above, for the sake of
simplicity, we focus on the non-local one, and we shall discuss the
local one in the conclusion.  As we have seen above, we study the
dynamics of the system after a {\it quench} from high temperature,
during which the system tries to decrease its energy to reach a low
temperature equilibrium state. In the following, we will focus only
on low temperatures such that the system has to encounter entropic
barriers during its relaxation process. 

In order to analyse the dynamics, we first map our model on a random
walker on the energy levels. It is on this model, which is much
simpler, that we get analytical results. In a second part, we check
numerically by Monte-Carlo simulations that the dynamics of the {\it
descent} model is quite well described by this random walker.  In
order to justify this mapping let us make some hypotheses.  We suppose
that the equilibration time in any energy level is short as compared
to the time spent to go through an entropic barrier between two energy
levels. That is to say we suppose that there is no entropic barrier
inside an energy level. This hypothesis is reasonable since the energy
levels are connected, and will be corroborated in the following by
numerical results. We also suppose that the paths between two energy
levels are uniformly distributed inside these levels. 

Following these hypotheses we compute the probability that the system
goes from one level to another using the mean features of the
levels. We take the slow dynamics of the model as equivalent to a
dynamics between the energy levels. In other words, we map the
dynamics of our model on a Markovian random walker on the energy
levels, whose features are precised in the following.  Since we know
the $D_n^k$ at low energy, and the number of paths starting from any
configuration, we know the total number of paths starting from one
level, $D_n^k(n-1)n/2$. We are searching the number of paths allowing
the system to decrease its energy from one level $k$. It is also the
number of paths starting from one level $l<k$ and arriving in $k$.
Since the energy variation cannot exceed $2$, $l \in \{k-2,k-1\}$.
Let us introduce a parameter $p(k)$ which represents the proportion of
paths starting from the level $k-1$ and allowing the system to {\it
increase} its energy by one unit.  It certainly depends on the energy
level.  We investigated numerically the configuration space of the
model, and we concluded that for $\frac{k}{n}=e < 0.1$, $p(k) \simeq
Ke$, with $K$ a constant that seems to depend very weakly on $n$. For
$n=10000$, we have $K \simeq 1.99$ and $K=1.97$ for $n=12$. For the
links between the fundamental state and the first excited ones
($k=1$), one can trivially show that $K$ exactly equals $2$. In the
following we will consider $K=2$.

\begin{figure}[h!]

\resizebox{!}{8cm}{
\includegraphics{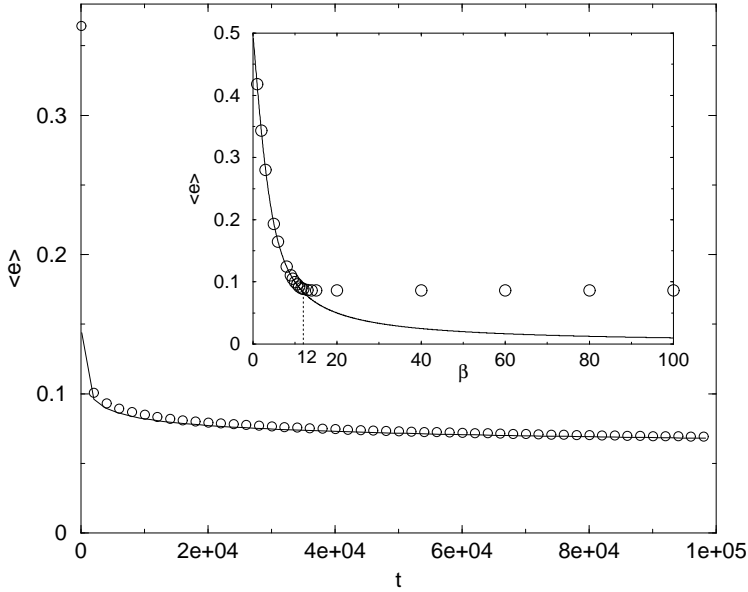}
}
\caption{
$\langle e\rangle(t)$ for $n=500$ and $T=0$. Circles represent
Monte-Carlo simulations of the {\it descent} model compared with the
evolution given by $1/(ln(Kt)+ln(ln(Kt)))$ (full line) which
represents the energy of the random walker with $p=Ke$, $K=2$. Circles in
the inset show $\langle e\rangle$ for the {\it descent} model at a
fixed time $t=10^4=\tau_{(\beta=12)}$ for various temperatures and the full line the
exact static energy. We check that for $\beta$ such that $\tau_{\beta} <
10^4$ the system has reached its equilibrium energy, and that for all $\beta$
such that $\tau_{\beta} > 10^4$ the system is out-of-equilibrium and its 
energy is independent of $\beta$.}
\label{f.1}
\end{figure}

So, $pD_n^{k-1}(n-1)n/2$ is the number of paths starting from one
level $k-1$ and {\it increasing} the energy by one unit, and
$pD_n^{k-1}/D_n^{k}$ the proportion of paths starting from one level
$k$ allowing the system to {\it decrease} its energy by one
unit. Being equilibrated in the level $k$, the system has to find one
of these paths to decrease its energy. These events are very rare
because of the great ratio $D_n^{k}/D_n^{k-1}$ at low energy. Indeed,
the typical time $\tau_{k,1}$, in Monte-Carlo steps, to go through the
entropic barrier (of height $\Delta S$) between the energy levels $k$
and $k-1$, satisfies:
\begin{equation}
\tau_{k,1}=\exp(\Delta S)=\frac{D_n^k}{pD_n^{k-1}} 
\simeq \frac{1}{p} \left(\frac{k+1}{k}\right)^n \stackrel{n \gg 1}{=} \frac{1}{p}\exp(n/k).
\end{equation}
This time gives us the probability rate by unit step that the random
walker goes from $k$ to $k-1$,
$\omega_{k\rightarrow k-1}=1/\tau_{k,1}$. In the same way we can
obtain the typical time $\tau_{k,2}\simeq\exp(2/(k/n-1/n))/q$ that the
system goes from a level $k$ to $k-2$, $q$ being the proportion of
paths starting from $k$ and arriving in $k+2$. Simulations show that $q$ is of order 1. 
We obtain $\omega_{k\rightarrow k-2}=1/\tau_{k,2}$,
but in the following we will always neglect it since
$\omega_{k\rightarrow k-2}/\omega_{k\rightarrow k-1}
\simeq \omega_{k\rightarrow k-1} \ll 1$. The master equation for the random
walker at $T=0$ is given by:
\begin{multline}\label{equ_WmasterT0}
P(k,t+1/n)=P(k,t)+\\
p(k+1)\exp(-n/(k+1))P(k+1,t)\\
-p(k)\exp(-n/k)P(k,t),
\end{multline}
where $t$ is in Monte-Carlo unit (one unit is equal to $n$ Monte-Carlo
steps).  By using the energy per particle $e$ and the continuous
limit in time and energy, this equation reads:
\begin{equation}\label{equ_difWT0}
\frac{\partial P(e,t)}{\partial t}=\frac{\partial}{\partial e}(p(e) P(e,t)\exp(-1/e)).
\end{equation}
Thus one obtains the evolution of the mean energy per particle
with time $\langle e\rangle(t)$, using $p(e)=2e$:
\begin{multline}
\frac{d \langle e\rangle}{d t}= \int_{0}^{(n-1)/n}\frac{\partial P(e,t)}{\partial t}e \, de \\
				= -2\int_{0}^{(n-1)/n}eP(e,t)\exp(-1/e)de.
\end{multline}
We calculate this integral by developing $eP(e,t)\exp(-1/e)$
around $\langle e\rangle$:
\begin{equation}\label{equ_difemT0}
\frac{d \langle e\rangle}{d t}=-2\e\exp(-1/\langle e\rangle)+\mathcal{O}(\Delta e^2 \exp(-1/\langle e\rangle)/\langle e\rangle^3),
\end{equation}
where one keeps only the zeroth order in $e-\langle e\rangle$, since
we shall demonstrate that $\Delta e^2=\langle (e-\langle
e\rangle)^2\rangle \simeq \langle e\rangle^2/2n$.  Indeed we can
derive time evolution of $\Delta e^2$ from eq.(\ref{equ_WmasterT0})
paying attention to keep terms of order $1/n$ since they cannot be
neglected in this case. As above, discrete sums are evaluated using an
integral and developing $eP(e,t)\exp(-1/e)$ around $\langle e\rangle$:
\begin{equation}\label{equ_de2}
\frac{d \Delta e^2}{d t}=-\frac{2}{\langle e\rangle}e^{-1/\langle e\rangle}\Delta e^2 +\frac{\e}{n}e^{-1/\langle e\rangle}-\frac{1}{n}e^{-2/\langle e\rangle}.
\end{equation}
The first term describes a rapid decay of $\Delta e^2$ towards its
quasi-equilibrium value $\langle e\rangle^2/2n$. Hence $\Delta e^2$ will be stabilized around $\langle e\rangle^2/2n$,
which is checked numerically for the {\it descent} model and the random
walker.  By integrating eq.~(\ref{equ_difemT0}) we obtain the time the
walker needs to go from high energy to the level $\epsilon=\langle e\rangle$,
$t=\epsilon\exp(1/\epsilon)/2+\mathcal{O}(\epsilon^2\exp(1/\epsilon))$ and:
\begin{equation}\label{equ_emtT0}
\langle e\rangle(t)\simeq \frac{1}{\ln(2t)+\ln(\ln(2t))}\stackrel{t\rightarrow \infty}{\sim}\frac{1}{\ln(t)}.
\end{equation}
This dynamics can also be found in a variety of theoretical models~\cite{Ritort_02}, such as
the {\it backgammon}
model and the {\it oscillator} model~\cite{oscillator}, 
as well as in compaction of granular media (models and experiments
\cite{r.deGennes_97,r.Knight_98,r.Knight_95}; see conclusion).
This result means that the system will never reach the fundamental
state at $T=0$, and will always stay out of equilibrium despite the
absence of energy barriers. Its relaxation time in the thermodynamical
limit is diverging. In the long time regime the dynamics will be
slower and slower. To check eq.(\ref{equ_emtT0}), we made some numerical
simulations on the random walker and on the original model. The
interest of the random walker is that its simplicity allows us to
compute numerically its {\it exact} dynamics for large $n$ and $t$ (up
to $t=10^{12}$) using its transition matrix. We find very good
agreement between analytical and numerical calculations. Monte-Carlo
simulations on the descent model are also in very good agreement with
analytical calculations at large $t$, as we can see in
fig.~\ref{f.1}. These results corroborate {\it a posteriori} the
hypothesis that there are no entropic barriers inside the energy
levels, since otherwise the random walker would be faster than the
{\it descent} model.

Let us remark that the only {\it analytically unknown parameter} is
$p(e)$.  In fact, this parameter is irrelevant for the asymptotic
dynamics while $p$ is of polynomial type: $p=Ke^{\alpha}$. Indeed,
using this equation for $p$, eq.~(\ref{equ_difemT0}) reads:
\begin{equation}
 \frac{d \langle e\rangle}{d t}=-K\e^{\alpha}\exp(-1/\langle
e\rangle)+\mathcal{O} (\Delta e^2 \exp(-1/\langle e\rangle)\langle
e\rangle^{\alpha-4}),
\end{equation}
and hence the mean energy per particle becomes:
\begin{equation}
\e(t)\simeq
\frac{1}{\ln(Kt)+(2-\alpha)\ln(\ln(Kt))}\stackrel{t\rightarrow
\infty}{\sim}\frac{1}{\ln(t)}.
\end{equation}
For $\alpha=2$, the evolution in $1/\ln(t)$ is even exact at all times
in the thermodynamical limit, not only asymptotically, which is well
verified in numerical simulations.

The variance $\Delta e^2$ does not depend on $\alpha$, since
eq.~(\ref{equ_de2}) becomes:
\begin{equation}
\frac{d \Delta e^2}{d t}=-2\e^{\alpha-2}e^{-1/\langle e\rangle}\Delta
e^2 +\frac{\e^{\alpha}}{n}e^{-1/\langle e\rangle}.
\end{equation}
We still have the quasi-equilibrium value $\Delta e^2= \e^2/2n$. We
will see below that the shape of $p$ does not change the aging in the
energy-energy correlation function either. It is not surprising: what
dominates the dynamics is the exponential decay of the random walker
transition rate with the inverse energy.

In the following we will consider again $p=2e$, since
it mimics the best the {\it descent} model. 

To study the dynamics at $T>0$, we have to consider the possibility
that the walker increases its energy. From the definitions of the
dynamics and of $p$ and $q$, it follows that $\omega_{k\rightarrow
k+1}=p\exp(-\beta)$ and $\omega_{k\rightarrow k+2}=q\exp(-2\beta)$,
where $\beta$ is the inverse temperature. At low temperature,
$\omega_{k\rightarrow k+1}/\omega_{k\rightarrow k+2} \ll 1$, so we
need not consider the possibility that the system jumps two levels in
one step. In the same way as at $T=0$, the evolution of the random
walker position follows a master equation, which in the continuous
limit, leads to:
\begin{equation}\label{equ_difemT}
\begin{split}
\frac{d \langle e\rangle}{d t}=-2\e&\exp(-\frac{1}{\langle e\rangle})
		+2\e\exp(-\beta)\\
&+\mathcal{O}(\Delta e^2\exp(-1/\langle e\rangle)/\langle e\rangle^4).
\end{split}
\end{equation}
We remark that $d \langle e\rangle/d t =0$ for $\langle
e\rangle=1/\beta=T$, which is the equilibrium energy for the {\it
descent} model for $T\ll 1$. At high $T$ the random walker
does not represent the {\it descent} model anymore since it evolves
in a region where $D_n^k \ne (k+1)^n$. Thus we focus on
the low $T$ region.  One can see from eq.~(\ref{equ_difemT}), that
while the system is far from its equilibrium energy the
possibility that it jumps to an higher level by thermal activation
will be negligible in front of its possibility to decrease its energy:
for $\langle e\rangle\gg 1/\beta$ we have $\exp(-\frac{1}{\langle
e\rangle})
\gg \exp(-\beta)$.  Far from its equilibrium
state, the system is not influenced by the temperature and its dynamics is
the same as at $T=0$. This defines a typical time
$\tau_{\beta}=\exp(\beta)/(2\beta)$ needed by the system to reach
$\langle e\rangle=1/\beta$ with the dynamics at $T=0$. Near equilibrium,
temperature effects appear and we obtain the dynamics by developing
$\exp(-1/\langle e\rangle)$ around $1/\beta$ in eq.~(\ref{equ_difemT}). We find an
exponential relaxation of energy with the same characteristic time
$\tau_{\beta}$ (see fig.~\ref{f.1}, inset). 

\begin{figure}[h!]
\resizebox{!}{8cm}{
\includegraphics{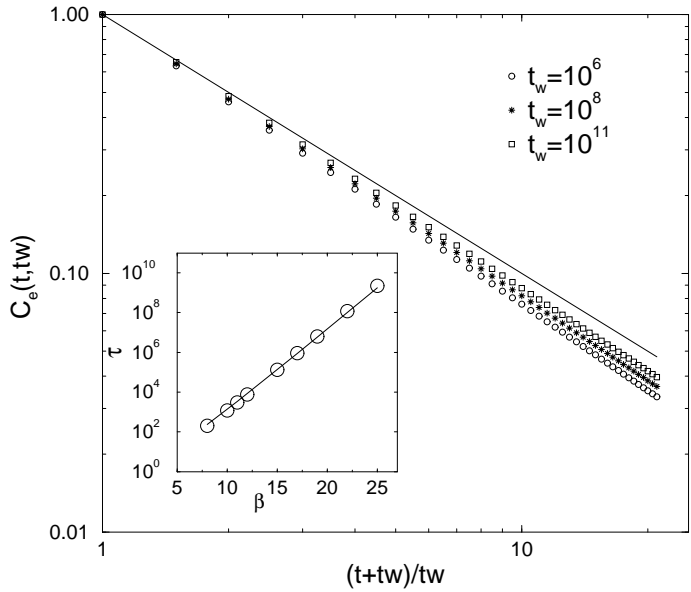}
}
\caption{
Energy-energy correlation of the random walker at $T=0$, for $n=2000$, 
as a function of $(t+t_w)/t_w$, compared with $t_w/(t+t_w)$ (full line).
The inset shows the relaxation time as a function of $\beta$, 
of the random walker, obtained from $C_e(t,t_w)$ at equilibrium, and the full line
a fit by $A\exp(\beta)/2\beta=A\tau_{\beta}$, with $A=1.25$, corroborating analytical calculations. 
}
\label{f.2}
\end{figure}

We define two types of correlation functions in order to characterize
the dynamics.  One is based on the matching of two permutations, in
others words on the proportion of sites at time $t$ which bear the
same particles as at a time $t_w$: $C_{\sigma}(t,t_w)=\sum_1^n
\delta_{\sigma_i(t+t_w),\sigma_i(t_w)}/n$, where $t_w$ is the waiting
time after the quench.  At high temperature, it can be proven that
$C_{\sigma}(t\gg t_w,t_w)=1/n$, at lower temperature we numerically
observe that it is also the case. This correlation function shows
rapid decay for any $t_w$, with no aging effects. The second
correlation function we consider is the energy-energy correlation:
\begin{equation}
C_e(t,t_w)=\frac{\langle e(t+t_w)e(t_w)\rangle-\langle
e(t+t_w)\rangle\langle e(t_w)\rangle}{\Delta e^2(t_w)}. 
\end{equation}
We studied it numerically for
both models at different $t_w$. For the random walker, at $T=0$ one can
see in fig.~\ref{f.2} that the system ages. The correlation function
seems to tend towards the scaling law: $C_e(t,t_w) \simeq t_w/(t+t_w)$ at very
large $t_w$. 

At $T>0$ we expect the dynamics to be the same as
at $T=0$ while the system is out-of-equilibrium. As long as $t <
\tau_{\beta}$ we find numerically the same plots as at $T=0$
whereas we find time translational invariance for $t+t_w >
\tau_{\beta}$. The same results are found for the {\it descent} model but
with statistical noise since we cannot use the transition matrix. 
These results show the strong importance of the observable we
focus on to observe aging \cite{r.pmadrid_02}.

\begin{figure}[h!]
\resizebox{!}{8cm}{
\includegraphics{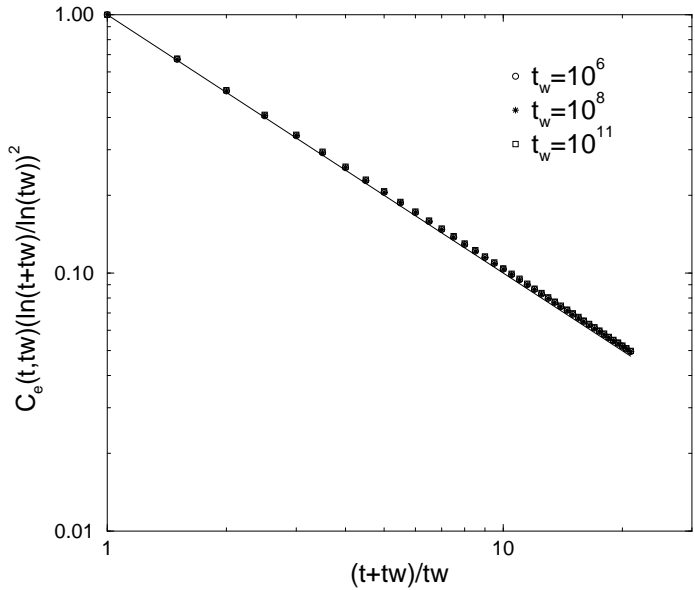}
}
\caption{ Energy-energy correlation corrected by
$\ln^2(t+t_w)/\ln^2(t_w)$ of the random walker at $T=0$, for $n=2000$,
as a function of $(t+t_w)/t_w$, compared with $t_w/(t+t_w)$ (full
line). The logarithmic corrections ensure the collapse of the data
for different values of $t_w$. }
\label{f.3}
\end{figure}

Now we propose a simple argument leading to the following 
law at large $t$ and $t_w$~:
\begin{equation}
C_e(t,t_w) \simeq \frac{t_w}{t+t_w} \; \frac{\ln^2 t_w}{\ln^2 (t+t_w)}. 
\label{correl}
\end{equation}
As illustrated in figure \ref{f.3}, this law accounts quite well for
our numerical observations as far as the random walker is
concerned. In particular, the logarithmic corrections ensure the
collapse of the curves for different values of $t_w$ on a straight
line of slope very close to -1. For the {\it descent} model, the
statistical noise does not allow to distinguish if the 
logarithmic corrections improve the agreement between theory 
and numerical experiments.

Our argument is as follows: since the dynamics on energy levels is
Markovian, the future of the random walker beyond $t_w$
({\em i.e.} $t>0$) only depends on the energy distribution at time $t_w$. 
For the {\it descent} model it will be only true at large times, when
the entropic barriers make the dynamics on the energy levels Markovian. 

Now we make the following hypothesis: the evolution of the
mean energy $\langle e \rangle(t+t_w)$ only depends on the 
mean energy $\langle e \rangle(t_w)$ and not on the precise 
shape and width of the energy distribution at $t_w$. This
hypothesis is supported by numerical simulations with 
different initial energy distributions and it is
corroborated by the excellent agreement between theoretical
and numerical correlation functions.  Furthermore, it amounts
to neglecting the terms of higher order in eq.~(\ref{equ_difemT0}),
as we made before, which is exact in the thermodynamics limit.

A consequence of this hypothesis is that any shift $\delta e$ at $t_w$ of the
mean energy has the same effect on $\langle e \rangle(t+t_w)$ as a time
delay $\delta t$ at $t_w$, since it only consists of a shift of the
initial condition. For small $\delta e$, $\delta t$
is such that:
\begin{equation}
\frac{\delta e}{\delta t} \simeq \frac{d \langle e \rangle}{dt}(t_w)
	= - \frac{1}{t_w \; \ln^2 t_w}.
\end{equation}
This point is illustrated in figure \ref{fig_aging}. Now we still denote by
$\langle e \rangle(t+t_w)$ the mean energy without any shift at $t_w$, and
by $\langle e | \delta e, t_w \rangle(t+t_w)$ the mean energy after an
energy shift $\delta e$ at $t_w$ (conditional mean). Subsequently
\begin{eqnarray}
\langle e | \delta e, t_w \rangle(t+t_w) & \simeq & \langle e 
        \rangle(t+ t_w  + \delta t) \\ \nonumber
& \simeq & \langle e \rangle(t+t_w) + \delta t \; 
	\frac{d \langle e \rangle}{dt}(t+t_w) \\ \nonumber
& \simeq & \langle e \rangle(t+t_w) + \delta e \; \frac{t_w}{t+t_w}
\; \frac{\ln^2 t_w}{\ln^2 (t+t_w)}.
\end{eqnarray}
Now the energy-energy correlation function is 
\begin{equation}
\langle e(t+t_w) e(t_w) \rangle = 
\sum_{e^*} \ \sum_e e \; e^* \; P(e^*,t_w) P(e,t+t_w|e^*,t_w)
\end{equation}
where $ P(e^*,t_w)$ is the probability that the energy per particle 
is $e^*$ at time $t_w$ and $P(e,t+t_w|e^*,t_w)$ is a conditional probability,
and 
\begin{eqnarray}
\lefteqn{\langle e(t+t_w) e(t_w) \rangle} \\ \nonumber
& = &
\sum_{e^*} e^* P(e^*,t_w) \langle e | \delta e = 
e^* - \langle e \rangle (t_w), t_w \rangle \\ \nonumber
 & = & \sum_{e^*} e^* P(e^*,t_w) \times \\ \nonumber
 & & \left[ \langle e \rangle (t+t_w) +
(e^* - \langle e \rangle (t_w)) \; \frac{t_w}{t+t_w} \; 
\frac{\ln^2 t_w}{\ln^2 (t+t_w)} 
\right] \\ \nonumber
 & = & \langle e \rangle (t+t_w) \langle e \rangle (t_w) + \\ \nonumber
 & & \frac{t_w}{t+t_w} \frac{\ln^2 t_w}{\ln^2 (t+t_w)} \sum_{e^*} P(e^*,t_w) 
\left[ e^* - \langle e \rangle (t_w) \right]^2.
\end{eqnarray}
The last sum is the variance $\Delta e^2$ of the energy distribution
at time $t_w$. Hence we get eq.~(\ref{correl}). Note that the same
kind of argument can be used to derive the same energy-energy
correlation function for the Backgammon model and is also in excellent
agreement with numerical simulations that we perfeormed
independently. Note that our correlation function is not the same as
the one calculated in reference \cite{r.godreche_96}, which explains
that is does not have the same expression.  With $p=Ke^{\alpha}$, this
law is unchanged as we saw before for $\e(t)$.

\section{Conclusion}

We have focused so far on the non-local dynamics. Let us now discuss
briefly the local one. Its interest lies in the fact that it is truly
one-dimensional since it respects the one-dimensional character of the
model. We shall see that even if its analysis is complicated by the
existence of energetic barriers (which freeze the dynamics at $T=0$),
the qualitative conclusions are the same as for the non-local
case. Indeed, one shows numerically that below an energy per particle
$e_m(n) \approx 0.1$, which depends slowly on $n$, nearly all states
are local potential minima. Therefore at $T=0$, the system is always
stuck in these minima, and at $T>0$ it has to pass over energy
barriers at each energy level in order to lower its energy.  However,
these energy barriers are always of height 1, and one needs a time of
order $\exp(\beta)$ to pass them.  Therefore, the times needed to pass
over {\em energetic} barriers and {\em entropic} barriers have the
same order of magnitude and the system is not substentially slowed by
energetic barriers, at least at $T>0$.  More precisely, we have
measured typical times $\tau^{\prime}(n,T)$ by evaluating the constant
diffusion of particles at equilibrium, at small but finite
temperatures and we conclude that $\tau^{\prime}(n,T)$ is of dominant
order $n^2 \exp(\beta)$. As compared to the non-local characteristic
times, this dynamics is $n^2$ times slower than the non-local
one. Note that this prefactor $n^2$ is also present at high
temperatures where it can be proven rigorously that
$\tau^{\prime}(n,\infty) \sim n^2$ \cite{r.Wilson_01}. At high
temperature, the $n^2$ term is certainly due to the diffusion of
particles over the whole system. At smaller temperature, one can think
that it results from the diffusion of descents acting as walls between
ordered domains, which allows the system to explore energy levels.  As
far as aging is concerned, our numericals results remain compatible
with the law: $C_e(t,t_w)=t_w/(t+t_w)$.

A natural continuation of the present work will be to investigate into
deaper detail the relationship between our model and compaction of
granular media. Indeed, equation~(\ref{equ_difemT0}) also governs
the evolution of the density in simple models of compaction
(see~\cite{r.Knight_98} for example). On the other hand, in these  {\it mean-field} models based on
a free volume argument, as well as in the {\it one-dimensional} descent model, the slow
dynamics is due to the necessity of complex and long rearrangements of
particles to optimize the organization of the system.

To finish with, we mention that as soon as a model presents an effective Markovian dynamics
between energy levels due to entropic barriers, the present analysis can be applied to 
this model. For example, the Backgammon model can be tackled in such a way and 
one obtains the correct laws for the mean energy $\langle e \rangle (t)$ and the
correlation function $C_e(t,t_w)$. 

\begin{acknowledgement}
We thank R\'emy Mosseri and Alexandre Lef\`evre for fruitful discussions.
\end{acknowledgement}


\onecolumn
{
\begin{figure}[h!]

\resizebox{!}{8cm}{
\includegraphics{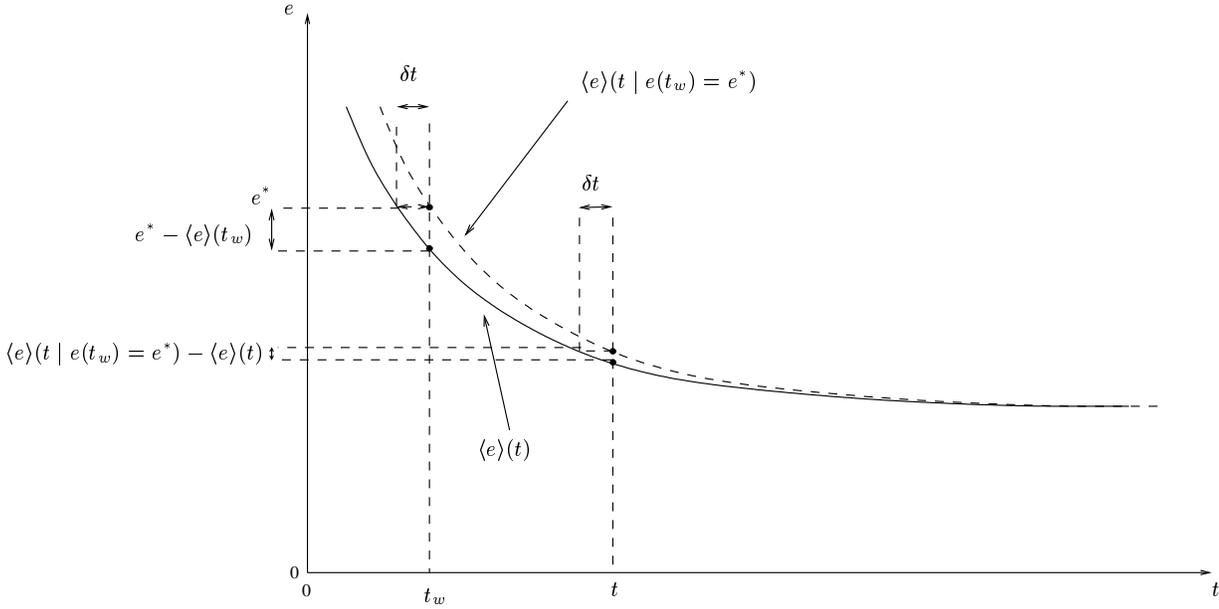}
}
\caption[]{Schematic picture of the aging analysis. The 
full line represents the mean energy per particle whereas the dashed line represents
the mean energy per particle when the system is at energy $e^{\ast}$ at time $t_w$.
$\delta t$ is the constant time shift between these two curves.}
\label{fig_aging}

\end{figure}
}


\end{document}